\documentclass[pdflatex,sn-mathphys-num]{sn-jnl}

\usepackage{graphicx}%
\usepackage{multirow}%
\usepackage{amsmath,amssymb,amsfonts}%
\usepackage{amsthm}%
\usepackage{mathrsfs}%
\usepackage[title]{appendix}%
\usepackage{xcolor}%
\usepackage{textcomp}%
\usepackage{manyfoot}%
\usepackage{booktabs}%
\usepackage{algorithm}%
\usepackage{algorithmicx}%
\usepackage{algpseudocode}%
\usepackage{listings}%

\theoremstyle{thmstyleone}%
\theoremstyle{thmstyletwo}%

\theoremstyle{thmstylethree}%

\begin{document}


\title{Searching for quantum-gravity footprint around stellar-mass black holes} 


\author*[1,2]{\fnm{Luigi} \sur{Foschini}}\email{luigi.foschini@inaf.it}

\author[3]{\fnm{Alberto} \sur{Vecchiato}}\email{alberto.vecchiato@inaf.it}
\equalcont{These authors contributed equally to this work.}

\author[4,5]{\fnm{Alfio} \sur{Bonanno}}\email{alfio.bonanno@inaf.it}
\equalcont{These authors contributed equally to this work.}

\affil*[1]{\orgdiv{Osservatorio Astronomico di Brera}, \orgname{Istituto Nazionale di Astrofisica (INAF)}, \orgaddress{\street{Via E. Bianchi, 46}, \city{Merate}, \postcode{23807}, \country{Italy}}}

\affil[2]{\orgdiv{Sezione di Milano}, \orgname{Istituto Nazionale di Fisica Nucleare (INFN)}, \orgaddress{\street{Via G. Celoria, 16}, \city{Milano}, \postcode{20133}, \country{Italy}}}

\affil[3]{\orgdiv{Osservatorio Astrofisico di Torino}, \orgname{Istituto Nazionale di Astrofisica (INAF)}, \orgaddress{\street{Via Osservatorio, 20}, \city{Pino Torinese}, \postcode{10025}, \country{Italy}}}

\affil[4]{\orgdiv{Osservatorio Astrofisico di Catania}, \orgname{Istituto Nazionale di Astrofisica (INAF)}, \orgaddress{\street{Via Santa Sofia, 78}, \city{Catania}, \postcode{95123}, \country{Italy}}}

\affil[5]{\orgdiv{Sezione di Catania}, \orgname{Istituto Nazionale di Fisica Nucleare (INFN)}, \orgaddress{\street{Via Santa Sofia, 64}, \city{Catania}, \postcode{95123}, \country{Italy}}}

\abstract{According to the asymptotically safe gravity, black holes may have characteristics different from those described according to general relativity if the running of the gravitational constant coupling happens at low energies. Particularly, they should be more compact, with a smaller event horizon, which in turn affects the other quantities dependent on it, like the photon ring and the size of the innermost stable circular orbit. We decided to test the latter hypothesis by searching in the literature for observational measurements of the inner radius of the accretion disk around stellar-mass black holes. We selected the smallest values measured when the disk was in high/soft state, made them homogeneous by taking into account the most recent and more reliable values of mass, spin, viewing angle, and distance from the Earth, and compared with the expectations of the Kerr metric. We do not find any significant deviation. Some doubtful cases can be easily understood as due to specific states of the object during the observation or instrumental biases. We set the tightest constraint on the parameter $\xi$ obtained to date.}


\maketitle

\section{Introduction}
The description of all physical phenomena in the Universe in one unified theory is the ultimate dream of theoretical physics. Despite of decades of theoretical research (see \cite{SPACETIME,ARMAS2021} for recent reviews), little has been obtained on the observational/experimental side. Any experimental or observational evidence of quantum gravity remains elusive still today \cite{ALVESBATISTA2023}. Today more than ever, it is necessary to find some observational constraint, to try stopping the uncontrolled proliferation of theories. As stressed by Eichhorn and Held \cite{EICHHORN2023}, ``irrespective of theoretical considerations, any observational avenue to put constraints on deviations from general relativity (GR), should be explored'' . This is also our philosophical approach to the problem, searching for \emph{any} deviation from GR around black holes\footnote{In the present work, we always adopted the Kerr metric as reference for GR calculations. Therefore, whether we talk about GR or explicitly write Kerr metric, we mean to refer to \cite{NOVIKOV1973}.}, although we eventually compare observations with expectations of asymptotic safe gravity (ASG), which we consider to be the best option presently at hands. 

First developed by Weinberg \cite{WEINBERG1979}, ASG has its pillars on quantum field theory and renormalization group (for example, see \cite{NIEDERMAIER2006,BONANNO2020} for reviews). The key point is that quantum effects are described by an unknown parameter $\xi$, which changes the gravitational constant, making it to depends on the energy scale. In the low energy limit (infrared), the running coupling can be written as:  

\begin{equation}
G(r) = G_{\rm N}\left(1-\frac{\xi}{r^2}\right)
\label{defG}
\end{equation}

\noindent where $G_{\rm N}= 6.67428\times 10^{-8}$~cm$^{3}$~g$^{-1}$~s$^{-2}$ is the classical gravitational constant.

ASG is expected to have observational consequences on astrophysical and cosmological scales \cite{REUTER2004}, and in particular around black holes \cite{REUTER2011,HAROON2018,EICHHORN2023,SANCHEZ2024} and neutron stars \cite{BONANNO2020B}. The main effect on black holes is a more compact object, although without curvature singularities \cite{EICHHORN2023}. A smaller event horizon affects also all the other related quantities such as the photon ring \cite{EICHHORN2023}, the ergosphere \cite{HAROON2018} and the radius of the innermost stable circular orbit $r_{\rm isco}$ \cite{SANCHEZ2024}. Particularly, S\'anchez \cite{SANCHEZ2024} suggested that differences up $\sim 30$\% in the $r_{\rm isco}$ are expected in the case of maximally rotating prograde black holes already with a small value of $\xi$ (see his Table~I, where he adopted $\tilde{\xi}=\xi/r_{\rm g}^2$, being $r_{\rm g}$ the gravitational radius of the black hole\footnote{Please note that S\'anchez \cite{SANCHEZ2024} adopted geometrized units with $c=G_{\rm N}=1$. Therefore, in his work $\tilde{\xi}=\xi/\bar{M}^2$, where $\bar{M}$ is the mass of the black hole in units of length, i.e. its gravitational radius.}). 

Some attempts to test these theories have been already made by using X-ray spectroscopy (see \cite{ZHANG2018,ZHOU2021,BAMBI2024}), but to date no deviations were found. We focus our study on the search for deviations from the value of the $r_{\rm isco}$ in the case of stellar-mass black holes calculated according to Kerr metric \cite{NOVIKOV1973}. Before starting new data analyses and model development, we decided to search in the scientific literature for cases worthy of study. 

Stellar-mass black holes are the smallest singularities, as primordial black holes were never observed to date. Therefore, the effects of ASG -- if any -- should be maximised. Moreover, since the peak of the emission from the accretion disk is inversely proportional to the square root of the mass of the compact object, it falls in the X-ray energy band, where there are less problems of contamination from the nearby environment. In the case of supermassive black holes, the disk emission peaks in the ultraviolet, where there are contaminations from emission lines of the broad-line region and the host galaxy. In the case of advection dominated accretion disks with low accretion rate, which peak at microwave/infrared frequencies \cite{ADAF}, it is possible to use the excellent spatial resolution of the Event Horizon Telescope to probe scales $\sim 10r_{\rm g}$ \cite{EHTM87,EHTSGRA}. However, $r_{\rm isco}$ can be very close to $r_{\rm g}$ in the case of maximally rotating prograde black hole. 

As a first step, we searched in all the available literature for measurements of the inner radius of the accretion disk when the latter was in high/soft state (see Sect.~\ref{data}). Then, we made all the values homogeneous so we can compare them with each other, and with the theoretical values calculated according to the Kerr metric. Given the measurement errors, we conclude by setting a lower limit to the $\tilde{\xi}$ parameter of a renormalization-group improved Kerr metric in the low-energy regime \cite{SANCHEZ2024}.

\section{A short summary of the accretion disk theory and models}
According to the standard theory \cite{SHAKURA1973,NOVIKOV1973}, the accretion disk around black holes is optically thick and geometrically thin, although there are other models, particularly in the case of very high or very low mass accretion rate \cite{ABRAMOWICZ2013}. From the observational point of view, the disk can be considered as a series of rings of increasing size going outward, each one emitting black body radiation with increasing temperature $T(r)$ as the radial distance $r$ from the black hole decreases (e.g. see Fig.~2 in \cite{EBISAWA2003}). The highest temperature $T_{\rm in}$ corresponds to the inner part of the disk, closer to the event horizon. The most used model in X-ray data analysis is the multicolor accretion disk \cite{MITSUDA1984,MAKISHIMA1986,MAKISHIMA2000}, and is implemented as \texttt{diskbb} model in the \texttt{xspec} software package\footnote{\url{https://heasarc.gsfc.nasa.gov/xanadu/xspec/}}. This model is based on a Newtonian approximation, but it can give results consistent with a Kerr black hole under certain inclination angles (small values, so that the Doppler boosting and the gravitational redshift are negligible) and with proper correction coefficients \cite{SHIMURA1995,KUBOTA1998}, as explained in many studies (see, for example, \cite{MERLONI2000,EBISAWA2003}). It is worth noting that {\tt diskbb} does not satisfy the zero-torque condition at the inner boundary. There are two options to cope with this issue: the most adopted and simplest option is to apply a correction factor \cite{KUBOTA1998} to the value of the inner radius of the accretion disk as measured by {\tt diskbb}, but there is also a specific {\tt xspec} model developed by \cite{ZIMMERMAN05}, namely {\tt ezdiskbb}, which gives similar results to the previous method since both are based on the same equation (cf Eq.~1 of \cite{KUBOTA1998} in the case of $\beta=1$, with Eq.~4 of \cite{ZIMMERMAN05}). 

There is also a fully relativistic model of multicolor accretion disk ({\tt kerrbb} in {\tt xspec}, \cite{KERRBB}). However, since we are searching in the literature, and {\tt diskbb} is the most adopted model across many years, we decided to refer to it for our calculations, and to apply the proper correction factors. In addition, when adopted, {\tt kerrbb} is used to estimate the spin, rather than $r_{\rm isco}$. Therefore, we can make some comparison only in those cases when the parameters $M$ (black hole mass), $i$ (viewing angle), and $d$ (distance), which are generally fixed in the model, are consistent with the values we adopted as reference (see Sect.~\ref{sec:cygx1}).

The {\tt diskbb} model can be summarised as follows. Although the emitted spectrum is a superposition of several blackbody at different temperatures, the disk luminosity $L_{\rm d}$ [erg~s$^{-1}$] can be calculated via the Stefan-Boltzmann law:

\begin{equation}
L_{\rm d} = 4\pi r_{\rm in}^2 \sigma T_{\rm in}^4
\label{disklum}
\end{equation}

\noindent where $\sigma=5.67\times 10^{-5}$~erg~cm$^{-2}$~s$^{-1}$~K$^{-4}$ is the Stefan-Boltzmann constant, and $T_{\rm in}$ [K] is the temperature at the inner radius $r_{\rm in}$ [cm]. This equation is valid when the outer radius of the accretion disk $r_{\rm out}$ is much larger than $r_{\rm in}$, so that the outer temperature $T_{\rm out}$ can be considered negligible \cite{MITSUDA1984,MAKISHIMA1986,MAKISHIMA2000}. 

However, X-ray spectra cannot measure the effective temperature, because of Comptonization and relativistic effects. The observed color temperature $T_{\rm col}$ is greater than the effective one by a factor $\kappa \sim 1.7-2.0$, which in turn depends on the accretion rate \cite{SHIMURA1995}. This is also linked to the effective location of the inner radius, which should be corrected by a factor $\varsigma=0.412$ \cite{KUBOTA1998}. Therefore, Eq.~(\ref{disklum}) is rearranged as:

\begin{equation}
L_{\rm d} = 4\pi\sigma \left(\frac{r_{\rm in}}{\varsigma}\right)^2 \left(\frac{T_{\rm col}}{\kappa}\right)^4
\label{disklum1}
\end{equation}

The luminosity can be calculated by means of geometrical considerations, since the thin disk can be assumed to be flat:

\begin{equation}
L_{\rm d} = 2\pi \frac{d^2}{\cos i} F_{\rm bol}
\label{disklum2}
\end{equation}

\noindent where $F_{\rm bol}$ is the disk bolometric flux [erg~cm$^{-2}$~s$^{-1}$], $d$ is the luminosity distance [cm], and $i$ is the viewing angle of the disk [deg]. By combining Eqs.~(\ref{disklum1}) and (\ref{disklum2}), it is then possible to estimate the inner radius of the accretion disk:

\begin{equation}
r_{\rm in} = \varsigma \kappa^2 d \sqrt{\frac{F_{\rm bol}}{2\cos i\sigma T_{\rm col}^4}}
\label{rin}
\end{equation}

The normalization $N$ of the \texttt{diskbb} model -- that can be directly estimated by spectral fitting -- includes some of the above factors, and is equal to:

\begin{equation}
N = \left(\frac{r_{\rm col}}{d_{10}}\right)^2 \cos i
\label{norm}
\end{equation}

\noindent where $d_{10}$ is the luminosity distance in units of $10$~kpc, and $r_{\rm col}=r_{\rm in}/\varsigma$ is the color radius. Therefore, the inner radius can also be calculated by rearranging Eq.~(\ref{norm}) as:

\begin{equation}
r_{\rm in} = \varsigma \kappa^2 r_{\rm col} = \varsigma \kappa^2 d_{10} \sqrt{\frac{N}{\cos i}}
\label{rinnorm}
\end{equation}

To conclude, as already noted and with all the caveats \cite{MERLONI2000,EBISAWA2003}, despite its simplicity, the \texttt{diskbb} model performs reasonably well, as it results from the comparison with more detailed relativistic models (e.g. \cite{EBISAWA1991,EBISAWA2003,FOSCHINI2006,LORENZIN2009}).

\section{Gathering data}
\label{data}
Eqs.~(\ref{rin}) and (\ref{rinnorm}) offer two possibilities to estimate the inner radius of the accretion disk. In addition to the quantities measured via the fit of the X-ray spectrum ($N$, $T_{\rm col}$, $F_{\rm bol}$), it is necessary to know the distance $d$, the viewing angle $i$, but also the mass of the black hole $M$ and its spin $a$ to estimate the radius of the innermost stable circular orbit $r_{\rm isco}$ according to GR (Kerr metric, \cite{NOVIKOV1973}): 

\begin{equation}
r_{\rm isco} = r_{\rm g} \left[3 + Z_{2} - \sqrt{(3-Z_2)(3+Z_1+2Z_2)} \right]
\label{risco}
\end{equation}

\noindent where $r_{\rm g}=G_{\rm N}M/c^2$ is the gravitational radius, $c$ is the speed of light in vacuum, $Z_1=1+\sqrt[3]{1-a^2}(\sqrt[3]{1+a}+\sqrt[3]{1-a})$, and $Z_2=\sqrt{3a^2+Z_1^2}$. This value will be the reference for comparison with observations. For example, in the case of a maximally rotating black hole ($a\sim 0.998$ according to Thorne \cite{THORNE1974}, $r_{\rm isco}\sim 1.24 r_{\rm g}$. Recently, Mummery \cite{MUMMERY25} revised the limit to the value $a\sim 0.99$, which implies $r_{\rm isco}\sim 1.45 r_{\rm g}$.

It is worth stressing, however, that the spin is the most critical value to be measured, and the debate is quite hot still today (e.g. \cite{FABIAN2012,REYNOLDS2008,SALVESEN2021,ZDZ2024A,ZDZ2024B,BEL2024,LASOTA2024}). There are basically two methods: the disk continuum fitting and the iron line modelling. The former is based on the determination of $r_{\rm isco}$, assuming that it is coincident with $r_{\rm in}$, by using {\tt diskbb} or any other model of accretion disk. Given $r_{\rm g}$ and by using Eq.~(\ref{risco}), then it is possible to calculate $a$. In practice, this is just the inverse of the method adopted in this work. Therefore, it is not suitable for our goal. 

The other method is based on the Fe~K$\alpha$ emission line profile, which is more distorted when the emitting matter is closer to the event horizon (for example, see figs. 3 and 5 in \cite{FABIAN2000}). Since the line is assumed to be emitted in the inner region of the accretion disk, this implies that $r_{\rm in}$ enters again in the modelling. However, while in the previous case the quantity to be measured is the peak temperature of the multicolor black body, now the observable is the Fe~K$\alpha$ line shape. This makes it suitable for our goal, since we are searching for deviations from GR. In both methods, the Comptonization plays a crucial role (e.g. \cite{ZDZ2024A,ZDZ2024B}): either by increasing $T_{\rm in}$ to the value of $T_{\rm col}$ in the multicolor black body, or by affecting the continuum at the basis of the Fe~K$\alpha$, which in turn can affect the effective shape of the line. Nonetheless, it seems that the two methods are consistent for maximally rotating black holes \cite{REYNOLDS2008,SALVESEN2021}. 

The values of $r_{\rm g}$, $a$, $d$, and $i$ changed often over time, depending on the technological evolution of the instruments and the observation conditions (e.g. finding the stellar companion is often difficult, given the presence of many molecular clouds and interstellar dust toward the Galactic centre). We have adopted the most recent values, which are not always consistent with those used by the authors we examined in the conversion of their observational data. For example, many authors did not report the normalization $N$ of the \texttt{diskbb} model, but only the calculated $r_{\rm in}$ or even $r_{\rm col}$. Therefore, to extract a value comparable with other observations made at different epochs and with different instruments, it is necessary to know which values of $d$ and $i$ were adopted in the conversion. This is not granted: sometimes the values used in the conversion were not written in the paper, which made it impossible to recalculate the original normalization. Other problems encountered when searching in the literature were: missing measurement errors, missing units of measurement, plain errors and/or typos. In case of errors, it was often possible to correct them by recalculating the described model. For example, when finding a weird normalization $N$, it is possible to recalculate it if $T_{\rm col}$ and $F_{\rm bol}$ were reported. Fluxes were often reported without measurement errors: in this case, a typical $10$\% error was adopted. 

\begin{figure}[h]
\centering 
\includegraphics[width=\textwidth]{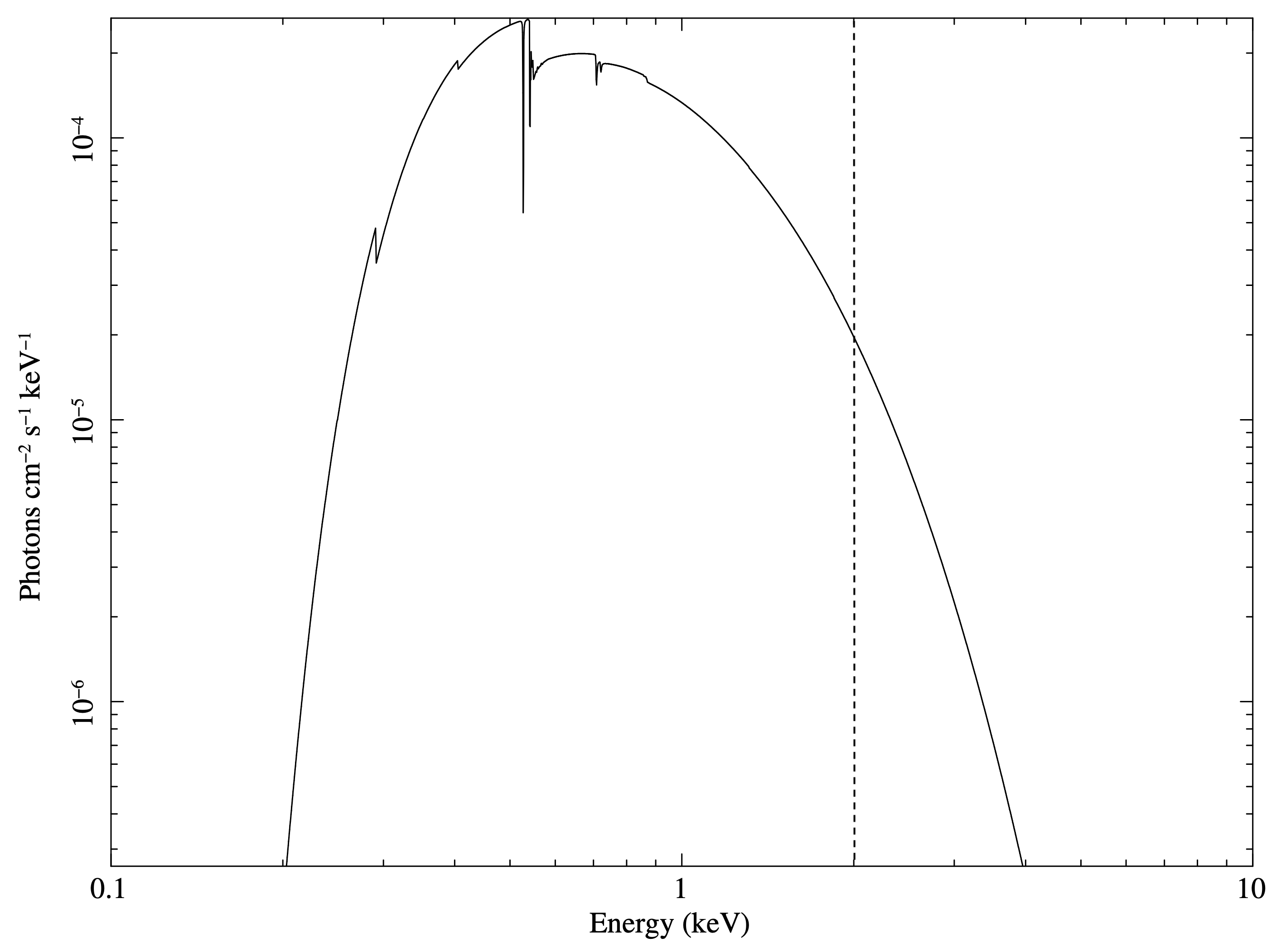}
\caption{\label{fig:threshold} Unfolded spectrum of the \texttt{diskbb} model with $T_{\rm col}=0.4$~keV, and assuming a typical Galactic hydrogen column $N_{\rm H}=10^{21}$~cm$^{-2}$. The vertical dashed line represents the $2$~keV low-energy threshold of \emph{RXTE}/PCA. The flux is in arbitrary units (\texttt{diskbb} normalization set to 1).}
\end{figure}

Yet another issue regards the instruments. In the case of stellar-mass black holes, the color temperature of the multicolor accretion disk peaks in the range $\sim 0.1-1$~keV. However, many detectors have a low-energy threshold of a few keV, which made it difficult to correctly extrapolate to lower energies. For example, one of the most used and prolific satellites in the observation of X-ray binaries was the \emph{Rossi X-ray Timing Explorer (RXTE)}\footnote{\url{https://heasarc.gsfc.nasa.gov/docs/xte/xtegof.html}}. The Proportional Counter Array (PCA) worked in the $2-60$~keV energy range, but its low-energy threshold degraded up to $3$~keV with the age. Already the $2$~keV threshold makes it difficult to correctly extrapolate a $T_{\rm col}=0.4$~keV, because only a small part of the spectrum can be detected, as shown in Fig.~\ref{fig:threshold}. 

In addition to the above cited problems, it is necessary to take into account that the accretion disk does not extend always down to the innermost stable circular orbit. Stellar-mass black holes change their activity from high/soft to low/hard states, and vice versa, passing through some intermediate states, depending on the accretion rate (e.g. \cite{ESIN1997,REMILLARD2006,BELLONI2010,MOTTA2021}). High/soft states occur at high accretion rates. The X-ray spectrum is dominated by the thermal emission of a geometrically-thin optically-thick accretion disk, which can extend down to $r_{\rm isco}$, and it accounts for more than $50$\% of the bolometric flux \cite{REMILLARD2006}. At low accretion rates (low/hard states), the thin disk is weak and cool, and a mildly relativistic radio jet is present. The X-ray spectrum is dominated by the optically-thin emission from the corona, modelled with a hard power law or a more detailed Comptonization model.  The intermediate state is an obvious mixture of the two above cited states. There are also many other variations, distinguished by small details\footnote{For example, see the plot: \url{https://www.issibern.ch/teams/proaccretion/Documents.html}}. However, for our purposes, we focus our attention on high/soft states, when the accretion disk extends downward the innermost stable circular orbit (i.e. $r_{\rm in}\sim r_{\rm isco}$). This is also a requirement to have a reliable measurement of $r_{\rm in}$, because the multicolor disk model tends to underestimate the inner radius when the corona significantly contributes to the bolometric flux \cite{MERLONI2000}, as it happens in the intermediate states, for example (see Sect.~\ref{sec:xtej1550}). 

When many measurements are available in the same article (X-ray spectra fit with different models, multiple observations because of a monitoring campaign), we selected the most promising cases, i.e. the best fit and the smallest values of $r_{\rm in}$. 

\section{Results}

\subsection{Cygnus X-1}
\label{sec:cygx1}
\begin{figure}[!ht]
\centering 
\includegraphics[width=\textwidth]{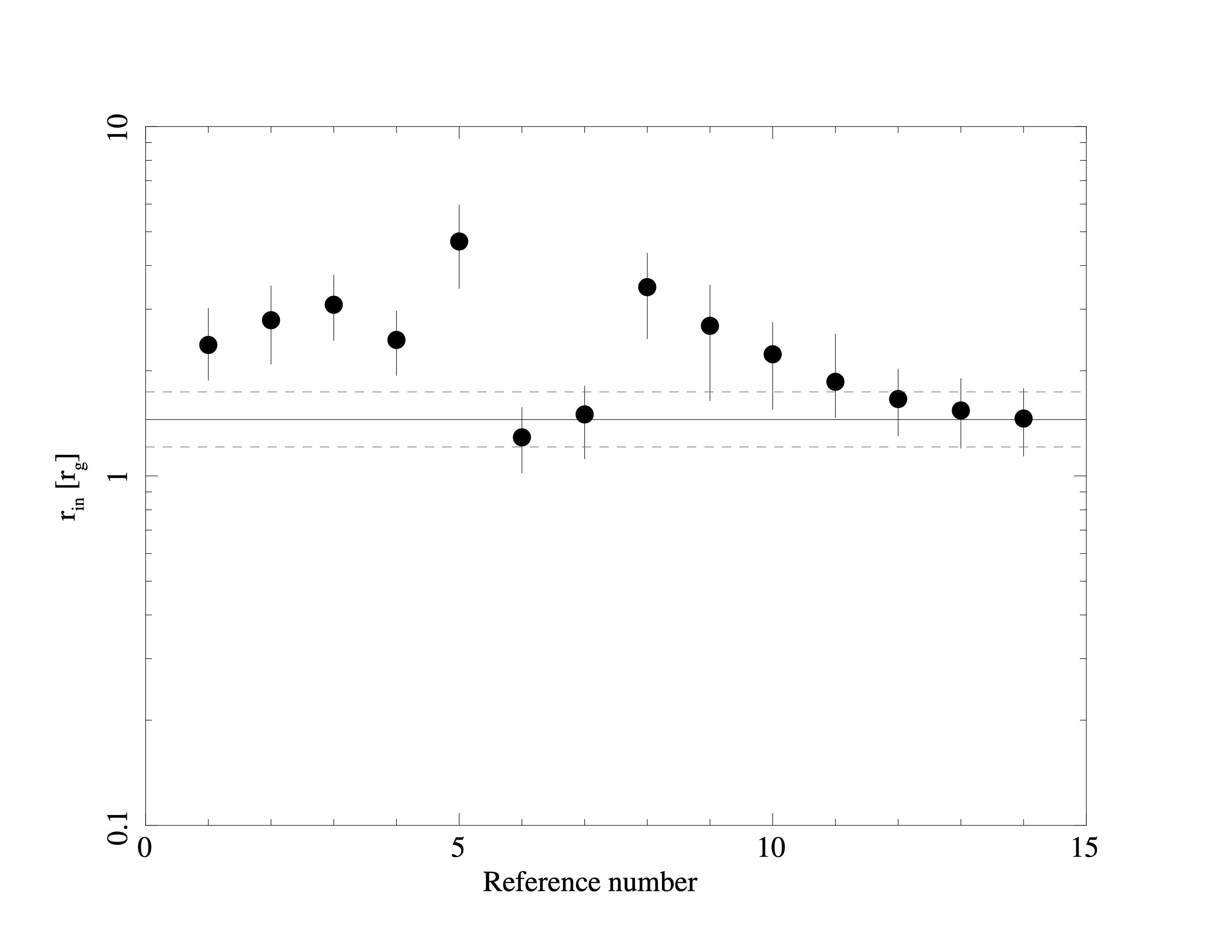}
\caption{\label{fig:cygx1} Cygnus X-1: inner radius of the accretion disk in units of [$r_{\rm g}$]. The two dashed grey lines represent the range of $r_{\rm isco}$ as expected from GR, while the continuous line represents the $r_{\rm isco}$ with the maximum allowable spin of $a=0.99$ according to \cite{MUMMERY25}. Reference number refers to the source of data: 1: \cite{DOTANI1997}; 2: \cite{POUTANEN1997}; 3-4: \cite{CUI1998}; 5: \cite{FRONTERA2001}; 6: \cite{TOMSICK2014}; 7: \cite{SUGIMOTO2016}; 8-11: \cite{WALTON2016}; 12-13 \cite{KUSHWAHA2021}; 14: \cite{YAN2021}.}
\end{figure}

This is the oldest and best known stellar-mass black hole. It has the most reliable measured values of $M=21.1_{-2.3}^{+2.2}M_{\odot}$, $d=2.22_{-0.17}^{+0.18}$~kpc, $i=27.51_{-0.57}^{+0.77}$~deg, and $a=0.9696-0.9985$ \cite{MILLERJ2021}. It is then possible to calculate $r_{\rm g}=31\pm 3$~km and $r_{\rm isco}=(1.21-1.74)r_{\rm g}$. The inner radius of the accretion disk in units of [$r_{\rm g}$] for different observations is displayed in Fig.~\ref{fig:cygx1}.  

The smallest value of $r_{\rm in}=(1.29_{-0.27}^{+0.28}) r_{\rm g}$ -- still consistent with the expectations from GR -- was derived from a simultaneous observation with \emph{Suzaku} and \emph{NuSTAR} ($1-300$~keV) between October 31 and November 1, 2012, by \cite{TOMSICK2014}. We considered the best fit model 4 in Table~2, where the normalization of \texttt{diskbb} is reported to be $N=20800_{-800}^{+1200}$, from which we can estimate $r_{\rm in}$ via Eq.~(\ref{rinnorm}). Unless explicitly stated otherwise, we always assumed $\kappa=1.7$. 

Kushwaha et al. \cite{KUSHWAHA2021} reported spectral fits with both {\tt diskbb} and {\tt kerrbb}. Therefore, it is possible to compare these two models. The latter has fixed the $M$, $i$, and $d$ to values similar to those by \cite{MILLERJ2021} and obtained a lower limit to the spin $a>0.9878$ for the Observation G ($r_{\rm isco}<1.49r_{\rm g}$), which is our number 12 in Fig.~\ref{fig:cygx1}, corresponding to $r_{\rm in}=(1.66\pm 0.36)r_{\rm g}$. Kushwaha et al. \cite{KUSHWAHA2021} reported also another measurement for the Observation I ($a>0.9843$, $r_{\rm isco}<1.55r_{\rm g}$), which we did not include in Fig.~\ref{fig:cygx1} as {\tt diskbb} gives a values greater than Obs G ($r_{\rm in}=1.89_{-0.41}^{+0.40}r_{\rm g}$). The two values are consistent with each other and with the expected value from the Kerr metric. Other authors \cite{KAWANO2017,ZDZ2024A,ZDZ2024B} adopted the {\tt kerrbb} model to estimate the spin from high/soft state observations, when $r_{\rm in}\sim r_{\rm isco}$, and they obtain values consistent with those of \cite{MILLERJ2021}.

\subsection{GRS~1915+105}
\begin{figure}[!ht]
\centering 
\includegraphics[width=\textwidth]{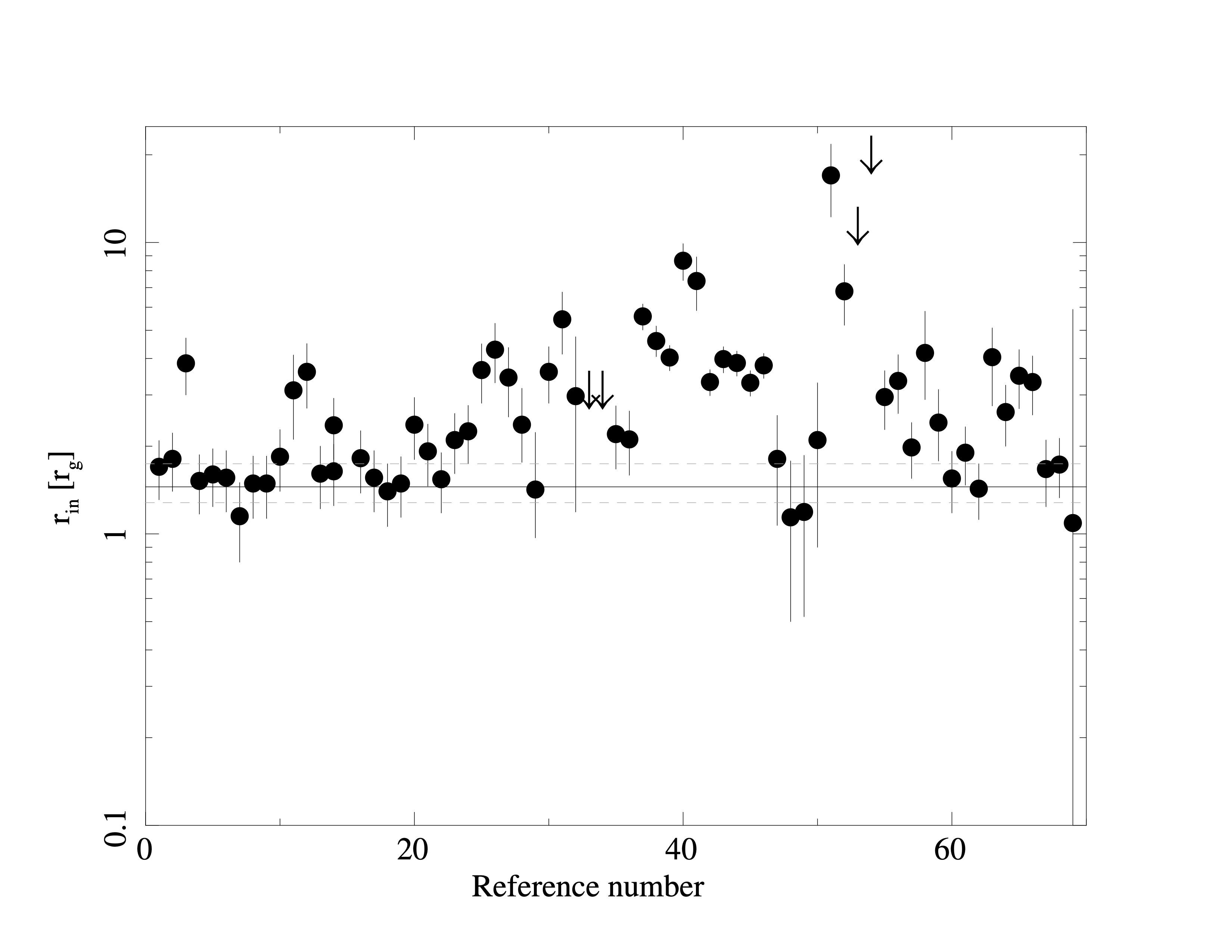}
\caption{\label{fig:grs1915} GRS~1915+105: inner radius of the accretion disk in units of [$r_{\rm g}$]. The two dashed grey lines represent the range of $r_{\rm isco}$ as expected from GR, while the continuous line represents the $r_{\rm isco}$ with the maximum allowable spin of $a=0.99$ according to \cite{MUMMERY25}. The arrows indicate upper limits. Reference number refers to the source of data: 1-2: \cite{TAAM1997}; 3: \cite{MUNO1999}; 4-7: \cite{FEROCI1999}; 8-10: \cite{RAO2000}; 11: \cite{BELLONI2000}; 12: \cite{ZDIARSKI2001}; 13-22: \cite{VADAWALE2001}; 23-27: \cite{UEDA2002}; 28-29: \cite{NAIK2002}; 30-32: \cite{DONE2004}; 33-36: \cite{OHKAWA2005}; 37-46: \cite{RODRIGUEZ2008}; 47-50: \cite{VIERDAYANTI2010}; 51-54: \cite{UEDA2010}; 55-60: \cite{RAHOUI2010}; 61: \cite{NEILSEN2011}; 62: \cite{MILLER2016}; 63-68: \cite{MINEO2017}; 69: \cite{HESS2018}.}
\end{figure}

This is one of the most monitored and studied stellar-mass black hole. Also in this case, there are reliable measured values of $M=11.8\pm 0.6M_{\odot}$, $d=9.4\pm 1.0$~kpc, $i=64\pm 4$~deg, and $a=0.970-0.997$ \cite{MILLER2013,SREEHARI2020,REID2023}. It is then possible to calculate $r_{\rm g}=17.4\pm 0.9$~km and $r_{\rm isco}=(1.28-1.74)r_{\rm g}$. The inner radius of the accretion disk in units of [$r_{\rm g}$] for different observations is displayed in Fig.~\ref{fig:grs1915}.

This is the only case where we adopted $\kappa=1.9$, as suggested by \citet{RODRIGUEZ2008} and \citet{NEILSEN2011}, since the color temperature is generally higher than usual for a stellar-mass black hole ($T_{\rm col}\sim 1-2$~keV). The values reported by \citet{RODRIGUEZ2008} -- indicated in Fig.~\ref{fig:grs1915} by the numbers from 37 to 46 -- are interesting, because they adopted the {\tt ezdiskbb} model. It was not possible to recalculate the normalization, because the inclination value is missing; therefore, we calculate $r_{\rm in}$ by using the peak temperature and the flux of the accretion disk from their Table~1. Obviously, in this case, we did not apply the $\varsigma$ correction factor by \cite{KUBOTA1998}, because the {\tt ezdiskbb} model already includes the zero-torque boundary condition.

Also for GRS~1915+105, $r_{\rm in}$ is consistent with the expectations from general relativity within the measurement errors. Some cases are borderline: one reason can be that the \texttt{diskbb} does not include relativistic corrections (gravitational redshift and Doppler boosting), which can become significant for high inclinations \cite{EBISAWA2003}. We remind that $i=64\pm 4$~deg for GRS~1915+105 \cite{REID2023}. 

\subsection{XTE~J1550-564}
\label{sec:xtej1550}
\begin{figure}[!ht]
\centering 
\includegraphics[width=\textwidth]{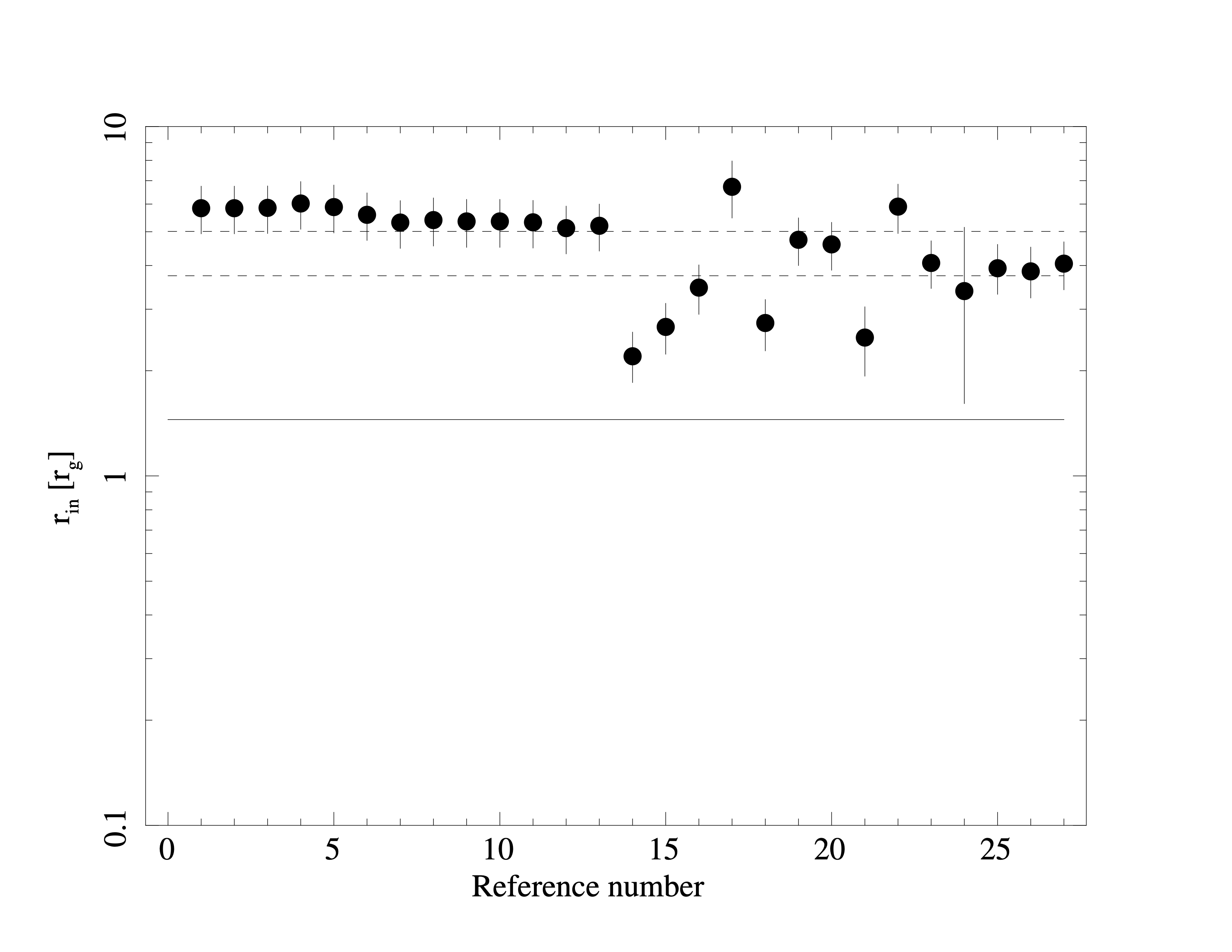}
\caption{\label{fig:xte1550} XTE~J1550-564: inner radius of the accretion disk in units of [$r_{\rm g}$]. The two dashed grey lines represent the range of $r_{\rm isco}$ as expected from GR, while the continuous line represents the $r_{\rm isco}$ with the maximum allowable spin of $a=0.99$ according to \cite{MUMMERY25}. Reference number refers to the source of data: 1: \cite{SOBCZAK1999}; 2-13: \cite{SOBCZAK2000}; 14-15: \cite{RODRIGUEZ2003}; 16: \cite{MILLER2003}; 17: \cite{KUBOTA2004A}; 18-20: \cite{KUBOTA2004B}; 21: \cite{SRIRAM2016}; 22: \cite{CONNORS2019}; 23-27: \cite{CONNORS2020}.}
\end{figure}

This black hole is interesting, because it is not maximally rotating. Its spin has been estimated to be $a=0.49_{-0.20}^{+0.13}$ \cite{STEINER2011}. The other quantities necessary to estimate the reference and measured values of the inner radii are $M=9.10\pm 0.61M_{\odot}$, $d=4.38_{-0.41}^{+0.58}$~kpc, and $i=74.7\pm 3.8$~deg \cite{OROSZ2011}. The gravitational radius is $r_{\rm g}=13.4\pm 0.9$~km, and $r_{\rm isco}=(3.74-5.01)r_{\rm g}$. The inner radius of the accretion disk in units of [$r_{\rm g}$] for different observations is displayed in Fig.~\ref{fig:xte1550}.

In this case, there are more significant deviations (reference numbers 14, 15, 18, 21). We should have removed these observations, but we thought they can be a good example of how the presence of Comptonization can alter the measurement of $r_{\rm in}$. The cases 14 and 15 were from \citet{RODRIGUEZ2003} and refer to RXTE observations in the $3-200$~keV energy band between April and June 2000. Table~1 of the cited paper reported the apparent radius calculated from the normalization of \texttt{diskbb} by assuming $d=6$~kpc and $i=73.1$~deg. The fits were done every one-two days and a thermal component is reported in 21 over 43 fits, with $T_{\rm col}\sim 0.37-0.95$~keV. We can already note that such temperature can be biased by the $3$~keV low-energy threshold of the detector. Since we are interested in searching for deviations from the expectations of general relativity, we selected the two smallest apparent radii, which are $r_{\rm col}\sim 32.5$~km (MJD $51662.2$), and $r_{\rm col}\sim 39.4$~km (MJD $51673.4$). There is also a third value of $r_{\rm col}\sim 39$~km at MJD $51682.3$, but the photon index of the power-law model is harder than those of the previous two cases ($1.76$ vs $2.33$, and $2.31$). This suggests the presence of a significant Comptonization, which can alter the estimate of the apparent radius \cite{MERLONI2000}, and therefore we did not consider it. For the two remaining cases, it is necessary to understand if they were measured during a high/soft state, with negligible corona contribution. As already written, the photon indexes of the power-law models were steep, but there are no indication of the fluxes due to the different components. We found some indications only in Fig.~4 in \citet{RODRIGUEZ2003}, which plots the fluxes in the $2-50$~keV energy band of the two models (power law vs \texttt{diskbb}). One point corresponding to MJD $51662.2$ is explicitly indicated: in this case, the power-law flux is about $2-3$ times that from the accretion disk. The other point, corresponding to MJD $51673.4$, is not indicated. However, by looking at the other points in the plot, one can reasonably think that the power-law component should have been at least comparable to the thermal emission. Therefore, we can conclude that these two inner radii estimates do not refer to high/soft states, and are affected by a strong Comptonization, which led to an underestimation of the value (cf \cite{MERLONI2000}). Similar cases are the number 18 from \citet{KUBOTA2004B} (it is the period 7 in Table 1), and the number 21 from \citet{SRIRAM2016} (it is the case C, model 2 in Table 2). 

A final note of warning: recently, the viewing angle has been challenged by \citet{CONNORS2019}. They suggested a value of $i\sim 40$~deg, much smaller than previously measured. This would have a significant impact on the estimation of the inner radius. The value calculated for $i\sim 75$~deg is $r_{\rm in}\sim 5.90r_{\rm g}$ (reference number 22), but in the case of $i\sim 40$~deg, then  $r_{\rm in}\sim 3.46r_{\rm g}$, smaller than expected from general relativity ($r_{\rm isco,min}\sim 3.74r_{\rm g}$). This new angle seems rather unlikely, because the kinematic analysis of the relativistic jet imposes a tight constraint on the viewing angle, indicating a value of $i\sim 71^{\circ}$ \cite{STEINER2012}. 

\begin{figure}[!ht]
\centering 
\includegraphics[width=\textwidth]{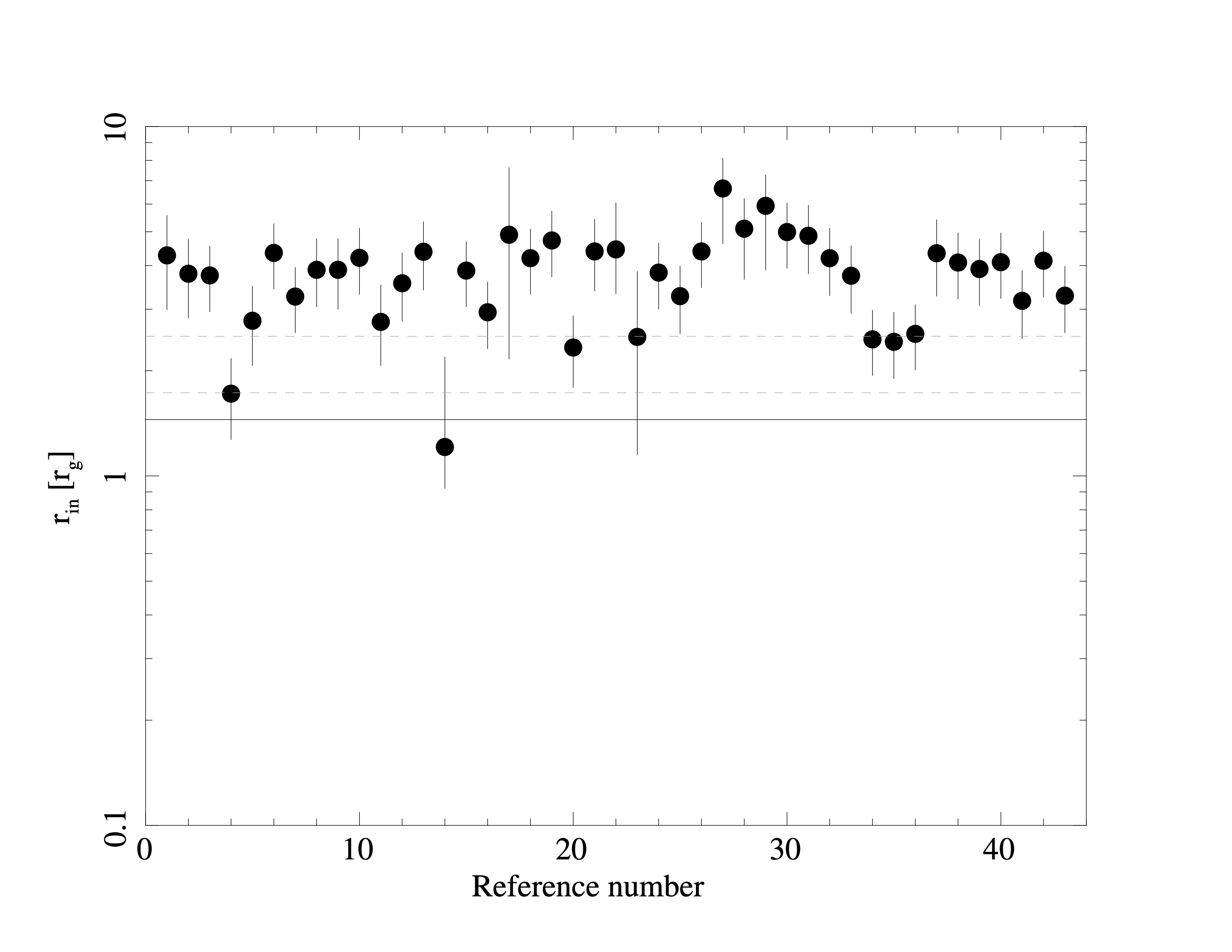}
\caption{\label{fig:gx339} GX~339-4: inner radius of the accretion disk in units of [$r_{\rm g}$]. The two dashed grey lines represent the range of $r_{\rm isco}$ as expected from GR, while the continuous line represents the $r_{\rm isco}$ with the maximum allowable spin of $a=0.99$ according to \cite{MUMMERY25}. Reference number refers to the source of data: 1-2: \cite{MILLER2004A}; 3: \cite{MILLER2004B}; 4-5: \cite{BELLONI2006}; 6: \cite{REIS2008}; 7: \cite{MILLER2008}; 8-10: \cite{DELSANTO2008}; 11-13: \cite{MOTTA2009}; 14-18: \cite{CABALLERO2009}; 19: \cite{SHIDATSU2011}; 20: \cite{MOTTA2011}; 21: \cite{TAMURA2012}; 22: \cite{RAHOUI2012}; 23: \cite{PLANT2014}; 24-25: \cite{LUDLAM2015}; 26: \cite{KUBOTA2016}; 27-29: \cite{STIELE2017}; 30-33: \cite{SRIDHAR2020}; 34-36: \cite{SHUI2021}; 38-40: \cite{YANG2023}; 41: \cite{PEIRANO2023}; 42: \cite{LIU2023}; 43: \cite{JANA2024}.}
\end{figure}

\subsection{GX~339-4}
Yet another well-known and observed stellar-mass black hole. Its reference quantities are: $M=9.0_{-1.2}^{+1.6} M_{\odot}$, $d=8.4\pm 0.9$~kpc, $i=30\pm 1$~deg, and $a=0.95_{-0.08}^{+0.02}$ \cite{PARKER2016}. The gravitational radius is $r_{\rm g}=13\pm 2$~km, and $r_{\rm isco}=(1.73-2.51)r_{\rm g}$. The inner radius of the accretion disk in units of [$r_{\rm g}$] for different observations is displayed in Fig.~\ref{fig:gx339}.

Also in this case, there are no significant deviations. The only value outside the expected range -- but stil consistent within the measurement errors -- is the number 14 \cite{CABALLERO2009}. It refers to a \emph{INTEGRAL} and \emph{XMM-Newton} observation ($0.7-200$~keV) performed between January and March 2007. The value of $r_{\rm in}=1.21_{-0.29}^{+0.98}r_{\rm g}$ has been estimated from the data of Epoch~1 of Table~2. We noted the presence also of a hard component modelled with a power law with $\Gamma\sim 1.46$, and Table~3 stated that the disk component accounts for $\sim 21$\% of the total flux. Again, a significant Comptonization led to an underestimate of the inner radius of the disk.

\subsection{XTE~J1650-500}
As we wrote in the Introduction, we expect the most significant deviations from general relativity to occur around the smallest black holes. Therefore, we decided to also study two objects with the smallest masses, although with large uncertainties and the lack of some reference information. The first one is XTE~J$1650-500$ and has its reference quantities are: $M=4.0\pm 0.6 M_{\odot}$, $d=2.6\pm 0.7$~kpc, $i=70\pm 4$~deg, and $a\sim 0.9982$ \cite{OROSZ2004,HOMAN2006,SLANY2008}. The gravitational radius is $r_{\rm g}=5.9\pm 0.9$~km, and $r_{\rm isco}\sim 1.23r_{\rm g}$. 

We found only two estimates in the literature, both showing no deviations from the expected value of $r_{\rm isco}$: $r_{\rm in}=(18\pm8) r_{\rm g}$ \cite{MILLER2002}, and $r_{\rm in}=(5.3\pm 1.7)r_{\rm g}$ \cite{MINIUTTI2004}. 

Searching for some unpublished data in the archives, we found one \emph{Swfit} observation (id 00031762001) performed on July 17, 2010. We analysed the data by using \texttt{HEASoft v. 6.33.2} software package and \texttt{CALDB} updated on August 28, 2024. \emph{Swift}/XRT data were reduced by using standard procedures (\texttt{xrtpipeline} with default values). We found no sources, with a flux upper limit of $\sim 3\times 10^{-11}$~erg~cm$^{-2}$~s$^{-1}$ ($3\sigma$ detection, $\sim 1.0$~ks exposure in photon counting mode, $N_{\rm H}=3.68\times 10^{21}$~cm$^{-2}$ \cite{NH}, power-law model with $\Gamma=2$). This means that XTE~J1650-500 was likely in a quiescent state. 

\subsection{GRO~J0422+32}
The latest object is a candidate for the smallest black hole known, with $M=2.7_{-0.5}^{+0.7} M_{\odot}$, $d=2.49\pm 0.30$~kpc, $i=55.6\pm 4.1$~deg \cite{GELINO2003,CASARES2022}. No information about the spin was found, and therefore it is not possible to estimate the expected $r_{\rm isco}$. We found only one paper useful to estimate the inner radius: $r_{\rm in}=(5.1\pm 2.3)r_{\rm g}$ \cite{SHRADER1997}. We also found one \emph{Swift} observation (id 00032976001) performed on October 3, 2013. We analyzed the data with the same procedure adopted for the above case, but we did not find any source, with an upper limit of the flux of $\sim 1.5\times 10^{-11}$~erg~cm$^{-2}$~s$^{-1}$ ($3\sigma$ detection, $\sim 1.2$~ks exposure in photon counting mode, $N_{\rm H}=1.51\times 10^{21}$~cm$^{-2}$ \cite{NH}, power-law model with $\Gamma=2$). Also GRO~J0422+32 was likely in a quiescent state. 

\section{Conclusions}
We searched in all the available literature for observations of X-ray binaries in high/soft state, where it is expected that the inner disk is closest to the innermost stable circular orbit ($r_{\rm in}\sim r_{\rm isco}$) according to the Kerr metric, to test if there are deviations from general relativity. We selected the measurements made by using the {\tt diskbb} model, we applied the best corrections for the Comptonization \cite{SHIMURA1995} and the inner boundary \cite{KUBOTA1998}, we made all the values homogeneous according to the most recent and reliable values of the reference quantities ($r_{\rm g}$, $a$, $d$, $i$). All measured radii are consistent with the expectations of general relativity. Some anomalous cases can be easily reconciled by taking into account Comptonization and how it affects the estimate of $T_{\rm in}$, uncertainties in the reference quantities, and instrumental biases.

Given these negative results, we would like to set a constraint on the values expected from asymptotically safe gravity according to S\'anchez \cite{SANCHEZ2024}. The best case for setting a constraint on $\xi$ in Eq.~(\ref{defG}) is the measurement of Cygnus X-1 made by \citet{TOMSICK2014} with \emph{Suzaku} and \emph{NuSTAR} ($1-300$~keV) on October 31 - November 1, 2012. Cyg X-1 has a reliable data set of ($r_{\rm g}$, $a$, $d$, $i$)\footnote{The spin of Cyg~X-1 has been measured both with the FeK$\alpha$ line \cite{FABIAN2012} and the continuum fitting, resulting in consistent values \cite{SALVESEN2021}, and in agreement with the spin-orbit coupling \cite{MILLERJ2021}.} and the selected case offers an excellent X-ray statistics. By assuming $a=0.98$ (the arithmetic mean of the measured values), which implies an expected $r_{\rm isco}=1.61r_{\rm g}$ from GR (Kerr metric), and considering the measurement errors in the above cited observation ($-0.27r_{\rm g}$ at 90\% confidence level, which is $1.645\sigma$), we would have been able to measure a significant ($5\sigma$) deviation from GR, corresponding to $r_{\rm isco}\lesssim 0.79r_{\rm g}$(\footnote{Obviously, $r_{\rm g}$ here refers to the classical value, which is used as reference for the comparison.}), if and only if:

\begin{equation}
\tilde{\xi}=\frac{\xi}{r_{\rm g}^2}\gtrsim 0.24
\end{equation}

\noindent corresponding to $\xi>2.3\times 10^{12}$~cm$^2$. As expected, given the currently available instruments, a significant weakening of $G$ is needed to obtain a measurable reduction of $r_{\rm isco}$, according to the S\'anchez's \cite{SANCHEZ2024} theory. Nonetheless, this is the tightest constraint obtained to date. If {\it NewAthena}\footnote{\url{https://www.cosmos.esa.int/web/athena/home}} will be realized, it will be possible to improve the statistics significantly.

We note that a change in $G$ would affect also the measurement of the mass of the black hole. In the case of XRBs, the mass is measured via the Kepler's third law, which includes $G$ \cite{CASARES2014}. Therefore, we should revise the mass of Cyg X-1 by using Eq.~(\ref{defG}) in the mass function equation (cf Eq.~1 in \cite{CASARES2014}). However, the orbit of the binary system (the semimajor axis is $\sim 0.244\, \mathrm{AU}\sim 3.7\times 10^{7}$~km, \cite{MILLERJ2021}) is much greater than the size of $r_{\rm isco}=1.61r_{\rm g}\sim 50$~km, and, given a dependence on $r^2$ of $G(r)$ (cf Eq.~\ref{defG}), the effect is negligible and smaller than the present measurement error.  

We cannot set any constraint on negative values of $\xi$ with this method. In this case, $r_{\rm isco}$ from ASG should be greater than the value from GR. However, a greater $r_{\rm in}$ is expected from state transitions in stellar-mass black holes. Particularly, in the hard state, the inner disk is truncated and $r_{\rm in}$ is of the order of tens $r_{\rm g}$ (see, however, \cite{DATTA2024}).  

Other, more complex expressions of $G$ are currently under study \cite{ALFIO}, and will be the topic of another essay.

Before concluding, we would like to underline the need to improve the measurements of the reference quantities, particularly the spin, which are crucial to calculate the GR values to be compared with. It is also necessary to better understand how the Comptonization affects $T_{\rm in}$. Last, but not least, more observations on the smallest black holes are needed. 

\bmhead{Acknowledgements}
LF thanks Roberto Della Ceca, director of the Brera Astronomical Observatory -- INAF, for partially funding this research, and Daniele Malafarina for useful questions after the presentation of this work at the \emph{IV International FLAG workshop: the Quantum and the Gravity}, (Catania, Italy, 9-11 September 2024).

\end{document}